\documentclass[prl, twocolumn]{revtex4}

\usepackage{graphicx}

\newcommand{\eW}{\ensuremath{\epsilon_{\mathrm{W}}}}
\newcommand{\eR}{\ensuremath{\epsilon_{\mathrm{R}}}}
\newcommand{\chandra}{\ensuremath{\Omega^{\star}}}
\newcommand{\fig}[1]{Fig.~\ref{#1}}
\newcommand{\eqn}[1]{Eqn.~\ref{#1}}

\newcommand{\unit}[1]{\ensuremath{\mathrm{\: #1}}}
\newcommand{\uAmps}{\unit{\mu A}}

\newcommand{\sfig}[1]{Fig.~S#1~\cite{epaps}}
\newcommand{\smov}[1]{Movie~#1~\cite{epaps}}
\newcommand{\smovs}[1]{Movies~#1~\cite{epaps}}
\begin{document}

\title{Polygonal excitations of spinning and levitating droplets}

\author{R.J.A. Hill}
\author{L. Eaves}
\affiliation{School of Physics and Astronomy, University of Nottingham, Nottingham NG7 2RD, UK}
\email{richard.hill@nottingham.ac.uk}

\date{1 August 2008}

\begin{abstract}
 The shape of a weightless spinning liquid droplet is governed by the balance between the surface tension and centrifugal forces. The axisymmetric shape for slow rotation becomes unstable to a non-axisymmetric distortion above a critical angular velocity, beyond which the droplet progresses through a series of 2-lobed shapes. Theory predicts the existence of a family of 3- and 4-lobed equilibrium shapes at higher angular velocity. We investigate the formation of a triangular-shaped magnetically levitated water droplet, driven to rotate by the Lorentz force on an ionic current within the droplet. We also study equatorial traveling waves which give the droplet 3, 4 and 5-fold symmetry.
 \end{abstract}

%\keywords{magnetic levitation; liquid drop; weightless; rotation; drop; hydrodynamics; azimuthal waves; equilibrium shape; shape bifurcation; instability; non-linear dynamics; self-gravitating objects; Kerr black hole}
\maketitle

In experiments published between 1863 and 1865, Plateau devised a way to
study the behavior of liquids in the absence of gravity by suspending olive oil in a density-matched water/alcohol mixture~\cite{plateau}. In studies of a spinning oil droplet driven by a rotating shaft, he observed that the droplet shape, spherical at rest, became flattened at the poles as the droplet gained speed, whilst the equatorial diameter increased. Beyond a critical angular velocity, the droplet evolved into a spinning non-axisymmetric shape resembling a triaxial ellipsoid, which developed into a two-lobed shape with increasing speed. Plateau's experiments were inspired by the idea that the action of surface tension on the spinning droplet could model the influence of self-gravitation on the shape of spinning astronomical bodies~\cite{cohen}. It was later realized that the droplet model could provide insights into the behavior of atomic nuclei~\cite{gamow,cohen}. The analogies between the behavior of a liquid droplet and fundamental physics at both the astronomical and nuclear length scales has inspired some of the greatest mathematicians and physicists to study the instabilities of a rotating droplet (see \cite{cohen}). Such analogies still attract great interest, as demonstrated by recent studies of the equilibrium shapes of relativistic rotating stellar masses~\cite{cardoso,cardosoprl06} and of atomic nuclei~\cite{schunck}.

Although the simplicity of Plateau's technique for studying weightless fluids on Earth is attractive, comparison of the results with theory is complicated by shape deformations due to the viscous drag of the surrounding fluid. Here we avoid the problem of drag by using a strong magnetic field $B$ to diamagnetically levitate centimeter-sized spinning water droplets in air. We have developed a `liquid electric motor' technique to spin the droplet~\cite{epaps} and to realize a stable equilibrium droplet shape that has striking triangular symmetry in the plane of rotation.
\begin{figure}
 \includegraphics[width=8cm]{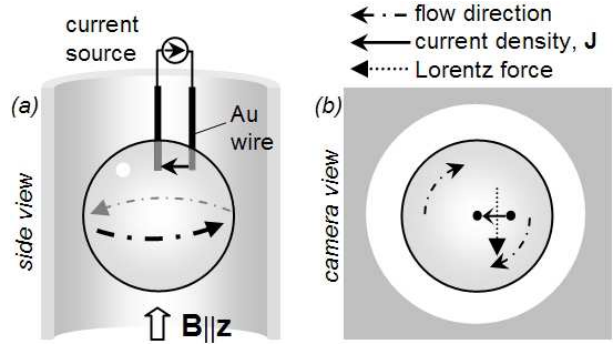}
 \caption{Schematic of a spinning, levitating droplet inside the magnet bore viewed from a) side and b) below magnet (camera view). The electrode positions, electric current and Lorentz force are indicated.}
   \label{f1}
\end{figure}

Brown and Scriven investigated the equilibrium shapes of a rigidly-rotating droplet theoretically using finite element analysis, grouping the shapes into `families' according to their symmetry~\cite{brown80}. The 2-lobed shape family, which includes the ellipsoid-like shapes observed in Plateau's experiments and 2-lobed `peanut' shapes observed in orbiting spacecraft~\cite{wang86,wang94} and in rolling `liquid marbles'~\cite{aussillous01}, branches from the axisymmetric family when the dimensionless angular velocity $\chandra = \sqrt{\rho\Omega^2 R^3/8\sigma}$ reaches 0.56. Here  $\Omega$, $\rho$ and $\sigma$ are the angular velocity, density and surface tension of the droplet respectively, and $R$ is the radius of the spherical droplet at rest~\cite{chandra65,brown80}. A 3-lobed and a 4-lobed family branches from the axisymmetric shapes at $\chandra=0.71$ and 0.75 respectively, but Brown and Scriven concluded that these shapes should be unstable to small shape perturbations and thus should be unobservable. However, experiments by Ohsaka and Trinh on acoustically levitated liquids showed that a $\sim 1$~mm-diameter droplet with a 3-lobed shape could be stabilized if it is forced into large periodic oscillation~\cite{ohsaka00}, but the droplet dynamics were far from equilibrium in their experiments.

Our liquid-motor technique generates a surface wave that travels around the droplet's equator in the opposite direction to the spin. By exciting a small-amplitude traveling wave with 3 nodes, we are able to suppress the degeneration of the triangular equilibrium shape (a member of the 3-lobed shape family) into the 2-lobed shape. This is similar to the stabilization technique used by Ohsaka and Trinh~\cite{ohsaka00}, but in our method the amplitude of the surface wave is small compared to the size of the droplet, so that the equilibrium shape is observed clearly. It is also possible to generate related large-amplitude traveling waves by our liquid-motor technique, and we demonstrate waves with up to 5 nodes in this paper.

We use a 16.5~T superconducting solenoid magnet with a vertical, 5~cm-diameter, room temperature bore to levitate a droplet of water~\cite{beaugnonNat, beaugnon91, beaugnon01, berry, sueda} with volume $V= 1.5$~ml, diameter $2R = 14 \unit{mm}$. We pass an electric current $I$ through the droplet by means of two thin, parallel, gold wire electrodes spaced $a=4.0\unit{mm}$ apart~\cite{epaps}. One wire is aligned with the vertical axis of the magnet bore and the droplet center, so that the ionic current flows perpendicular to $B$. The resulting Lorentz force generates a torque $\tau = BIa^2/2$ on the droplet which causes it to rotate around an axis parallel to $B$, through the center of the droplet (\fig{f1} and \sfig{1}). We add a small quantity of surfactant to reduce the surface tension of the liquid from $74 \unit{mJ\,m^{-2}}$ to $34\pm 2 \unit{mJ\,m^{-2}}$.

We increase $\tau$ gradually by increasing $I$ at a constant rate from $I=0$ to $I=600 \unit{\mu A}$ in 1~minute and then hold $I$ steady to maintain a constant $\tau$. The spinning droplet is axisymmetric, bulging at the equator and flattened at the poles. As it picks up speed, the equatorial radius increases.
\begin{figure}
 \includegraphics[width=8cm]{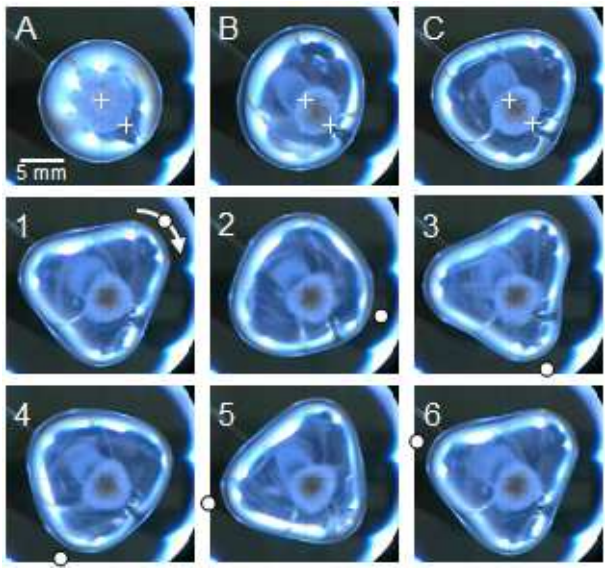}
 \caption
 {A 1.5~ml droplet levitating in the vertical magnet bore, viewed from below. A) not rotating. B) liquid rotating at $\Omega/2\pi\sim 2.0$~rps: the droplet's equator has an elliptical outline. C)$2.5$~rps: the equator has the symmetry of an equilateral triangle. In (B) and (C) the \emph{outline is not rotating} with the fluid. Crosses show the positions of the electrodes. 1-6) Six consecutive movie frames, 40~ms apart; here, the \emph{triangular outline is rotating} with the fluid, in the direction of the arrow, at $\Omega/2\pi=3.33$~rps. White circles follow one of the corners.}
 \label{f2}
\end{figure}
When $\Omega$ exceeds $\sim 2$~revolutions per second (rps), the equatorial outline becomes slightly elliptical (\fig{f2}B). This shape does not rotate with the flow of the liquid but remains fixed relative to the position of the off-axis electrode. Then, 20~s later, a 2-node oscillation develops, traveling azimuthally around the elliptical outline at $2.5 \pm 0.1 \unit{rps}$ (\smov{1}). However, less than 10~s after the current reaches $I=600 \unit{\mu A}$ the equatorial outline of the droplet develops the 3-fold symmetry of an equilateral triangle.  This outline remains fixed relative to the position of the off-axis electrode (\fig{f2}C and \smov{2}). Approximately 10~s later, a 3-node oscillation develops, moving around the equator, over the surface of the static shape, in the same direction as the flow of the liquid (\smov{3}). The amplitude of the oscillation increases during the next few seconds, whilst the amplitude of the lobes of the static shape on which it is superposed dies away. Once the static shape has decayed, approximately 5~s after the onset of the oscillation, the droplet has a regular triangular outline which rotates at $\Omega/2\pi=3.33 \pm 0.05 \unit{rps}$ (see \fig{f2}(1-6) and \smov{4}).  Close to the axial electrode the fluid is rotating faster, at $\sim 5 \unit{rps}$, as is evident from the vortex of small bubbles liberated by hydrolysis at the electrode, and there is additional vorticity generated at the off-axis electrode. Away from the electrodes however, the frequency with which tracer particles orbiting at a radius $\sim 0.3R$ from the axis perform a complete revolution around the droplet agrees well with the angular velocity of the droplet outline (\sfig{2}). This means that we can describe the rotation of the droplet as rigid body-like, in the sense that the fluid has a constant `background' vorticity $\xi=2\Omega$ (like that of rigid rotation) plus some localized vorticity generated by the electrodes.

Immediately after the onset of the triangular rotating shape we observe a decrease in the rotation rate of the fluid in the axial vortex, indicating a transfer of angular momentum from the vortex to the rest of the fluid (\sfig{2}). Similar behavior is observed at the onset of the 2-node oscillation during spin-up. The triangular rotating shape remains stable for several seconds longer ($\sim 100$ revolutions) before the rotation becomes eccentric (cam-like).

The equatorial shape of the droplet following the onset of the triangular shapes can be described by the relation
\begin{equation}
   r(\phi, t) = R[1+\eW \cos(m\phi) + \eR \cos(m(\phi-\Omega t))],
\label{e1}
\end{equation}
representing two superposed oscillations with the same wavenumber ($|m|=3$), similar maximum amplitudes, $\eW(\mathrm{max}) \approx 0.3$ and $\eR(\mathrm{max}) \approx 0.3$, but differing frequencies. Here $\phi$ is the azimuthal angle, increasing clockwise in \fig{f2}. The $\eW$-term represents the static triangular shape and the $\eR$-term represents the clockwise-rotating triangular rigid-body-rotation (RBR) shape. (\eqn{e1} also describes the static elliptical shape with superposed 2-node oscillation, observed during spin-up, if we set $m=2$.) In the frame of the rotating droplet, the static shapes are revealed to be azimuthally traveling waves, with frequency $\omega=m\Omega$, excited by the action of the off-axis electrode moving relative to the fluid. To clarify this point, we give the equation for the droplet outline in the \emph{rotating} frame,
\begin{equation}
   r(\phi', t) = R[1+\eW \cos(m(\phi'+\Omega t)) + \eR \cos(m\phi')],
\label{e2}
\end{equation}
where $\phi'=\phi-\Omega t$. Comparison of the $\eW$-term in \eqn{e1} with that in \eqn{e2} demonstrates that the static shapes observed in the laboratory reference frame are observed as traveling waves in the rotating frame. To avoid confusion we shall always use the phrase `static shape' to mean a shape that does not rotate in the laboratory reference frame (i.e. not rotating in the video images), and similarly for `rotating shape'. We emphasize that the static shape is generated from a surface wave moving relative to the fluid.

The frequency $\omega$ of the wave term depends on $\Omega$ through the effect of Coriolis and centrifugal forces~\cite{busse}. However, in our experiments the waves are `attached' to the off-axis electrode, so that a wave mode with $m$ nodes is excited only at a particular angular velocity $\Omega=\omega_m(\Omega)/m$. As the wave amplitude $\eW\rightarrow 0$, the kinematics of the droplet tend to that of RBR and we can compare our results with theory developed for equilibrium RBR shapes. Theory predicts a bifurcation point where 3-lobed equilibrium shapes may develop from the axisymmetric equilibrium shape when $\chandra$ reaches $\chandra_3=0.71$~\cite{brown80}. In our experiments, the triangular rotating shape appears at $\chandra =0.73 \pm 0.05$, in good agreement with theory. Although $\eW$ is small, so that the kinematics are close to RBR, it does not vanish completely ($\eW \sim 0.1 \eR$); the influence of the traveling wave is just discernable in \fig{f2} as a small periodic oscillation in the amplitude of the lobes of the rotating shape.  In this particular droplet, the condition $\omega=m\Omega$ for excitation of $m=3$ \emph{traveling waves} coincides with the theoretically predicted bifurcation point $\chandra \approx 0.71$. We observe the stable 3-lobed RBR shape only for a limited range of $R$ and $\sigma$. This suggests that, whilst theory shows that the equilibrium of 3-lobed RBR should be unstable to 2-lobed shapes, the 3-lobed RBR shape is stabilized by interaction with this small-amplitude traveling wave. In acoustic levitation experiments on smaller (10~$\mu$l) droplets, Ohsaka and Trinh~\cite{ohsaka00} obtained similar stabilization of a 3-lobed shape by exciting the axisymmetric $l=2, m=0$ spherical harmonic, but due to the relatively large amplitude of this oscillation, the kinematics in their experiments are far from RBR. The 3-lobed and 4-lobed rotating shapes have been observed in spinning droplets suspended in density-matched liquids, but in this case the shapes are deformed considerably by viscous drag~\cite{wangreview}. For comparison, we observe the onset of the 2-lobed RBR at $\chandra=0.57 \pm 0.05$, in good agreement with the theoretical value $\chandra_2 = 0.56$~\cite{chandra65,brown80}.

\begin{figure}
 \includegraphics[width=8cm]{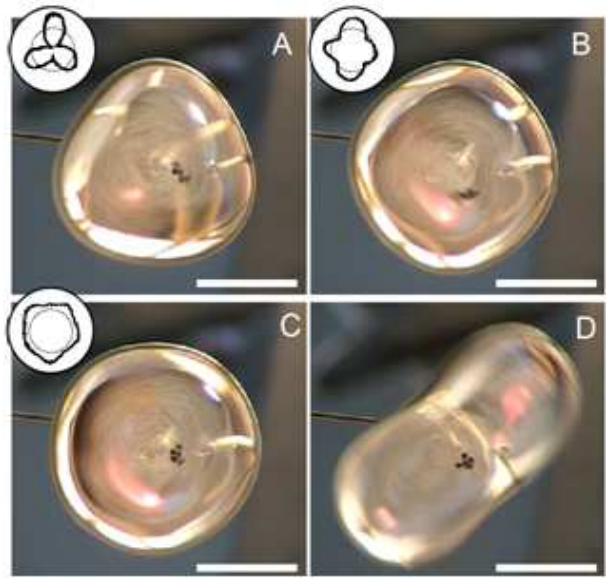}
 \caption
 {
    A-C) triangular, square and pentagonal shapes of a levitating 3.1~ml droplet. The fluid is rotating at $\sim 3$~rps but the outline remains fixed in the camera's view. Insets: polar plots of the amplitude of the equatorial bulges $\Delta r(\phi) = r(\phi)$ - $R$, where $r(\phi)$ is the equatorial radius and $\phi$ is the azimuthal angle. The radius of the circular frame represents 2~mm amplitude.  D) Clockwise rotating 2-lobed `peanut' shape. The scale bars represent 10~mm.
 }
 \label{f3}
\end{figure}

We now describe the excitation of azimuthally traveling waves with up to 5 nodes in a 3~ml ($2R = 18 \unit{mm}$) droplet, without surfactant added. Further examples are given in the supporting text~\cite{epaps}. We increase $I$ to 600~$\uAmps$ in 1~minute, then by increasing $I$ slowly, at $20 \uAmps /\unit{min}$, from this point, we can observe the 2-node oscillation evolve over approximately 2~minutes into a (nearly) rigidly rotating shape resembling a triaxial ellipsoid (\smovs{6-8}) with major axis perpendicular to the rotation axis and $\Omega/2\pi=2.6 \pm 0.1\unit{rps}$ ($\chandra=0.58 \pm 0.03$). When $I$ reaches $660 \uAmps$ the amplitude of the 2-lobed rotating shape decays and the droplet takes on the 3-fold symmetry of an equilateral triangle (\fig{f3}A and \smov{9}). When $I$ is kept constant at this point, the triangular static shape persists for over 3 minutes ($\sim 500$ revolutions) before an instability to strongly eccentric rotation causes the drop to break free of the electrodes. However, when we increase $I$ to $680 \uAmps$, the outline spontaneously develops a four-fold (square-shaped) symmetry; this shape is also fixed relative to the electrodes (\fig{f3}B and \smov{10}). The square shape persists for approximately 10~s ($\sim$ 30 revolutions), until the outline of the drop gains yet another corner, taking on the five-fold symmetry of a static regular pentagon (\fig{f3}C and \smov{11}). Approximately 10~s later, the symmetry of the pentagonal shape breaks spontaneously and degenerates to a 2-lobed `peanut' shape in under 2~s, with major axis length $3.0 \pm 0.2 \unit{cm}$, $\Omega/2\pi=1.70 \pm 0.04  \unit{rps}$ (\fig{f3}D and \smov{12}). The triangular, square and pentagonal shapes appear at $\Omega / 2\pi = 2.7 \pm 0.1$, $2.9 \pm 0.1$ and $3.1 \pm 0.2\unit{rps}$ respectively (\sfig{3}). These shapes are azimuthally-traveling waves in the reference frame of the rotating droplet, with $m=3, 4, 5$. Previously, traveling waves with $m=2$ and the related tesseral oscillation ($l=2, m=\pm1$) have been observed using acoustic levitation~\cite{trinh82, trinh88}. Busse~\cite{busse} has calculated $\omega$ for small-amplitude oscillations, treating small rotation rates as a perturbation and Radyakin~\cite{radyakin} investigated the frequencies of the $l=2, m=\pm 2, \pm 1, 0$ spherical harmonics at higher rotation rates using numerical methods, but the calculations for the $|m|=3$ modes have yet to be attempted. Ludu and Draayer have developed a non-linear theory of large-amplitude traveling waves, on the scale observed in our experiments, for non-rotating droplets~\cite{ludu}, but the case of rapid rotation has not been studied. For comparison, our measured critical frequencies are approximately $30\%$ smaller than those predicted by Busse.

Beaugnon et al. have already demonstrated the usefulness of magnetic leviation for studying the dynamics of weightless droplets~\cite{beaugnon01}. By combining magnetic levitation with a `liquid electric motor' spinning technique, we have observed the following striking features in the dynamics of a spinning droplet: i) a triangular rotating shape closely related to the 3-lobed rigidly rotating shapes considered theoretically, ii) large-amplitude azimuthally traveling waves with $|m|=2$ to 5.

The theoretical stability of the shapes of self-gravitating objects such as stars, planets and asteroids follows a similar pattern to that of spinning droplets~\cite{eriguchi82}: the axisymmetric shape (Maclaurin spheroid) of a self-gravitating body becomes unstable to the Jacobi (tri-axial ellipsoid) shape family at a critical angular velocity, just as the axisymmetric shapes of a water droplet becomes unstable to the 2-lobed shape family. At larger angular velocity other equilibrium shape families are predicted~\cite{eriguchi82}, such as the `triangle' sequence analogous to the 3-fold symmetry shape described here. Recently, studies of the light curves of Kuiper-belt objects have identified several rapidly spinning bodies that are likely to have a tri-axial shape due to their large angular momentum~\cite{lacerda}. There are also close analogies between the behavior of a liquid droplet and highly relativistic objects such as a black hole. The event horizon of the black hole can be thought of as a membrane endowed with surface tension~\cite{smarr}. In four dimensions, it is predicted that the axisymmetric shape of the rotating (Kerr) black hole remains stable up to the angular momentum limit imposed by the Kerr metric~\cite{cardosoprl06}. However, it has been proposed~\cite{cardoso,cardosoprl06} that in a higher dimensional space-time the horizon may become unstable to lower-symmetry shapes analogous to the non-axisymmetric shapes that we observe. The liquid droplet model of atomic nuclei is well known. Recently, the shape instabilities of a rapidly spinning liquid droplet has stimulated investigation of the Jacobi shape transition in rapidly rotating atomic nuclei, for which there is some experimental evidence (\cite{schunck} and references therein). For many years studying the behavior of a liquid droplet has proven to be a highly effective way to gain an intuitive understanding of the behavior of objects on much larger (astronomical) and smaller (nuclear) scales. The experimental results presented here should stimulate further insights.

\begin{acknowledgements}
We acknowledge M. Swift, A. Padilla and E. Copeland for useful discussions and EPSRC for support.
\end{acknowledgements}

%%%%%%%%%%%%%%%%%%%%%%%%%%%%%%%%%%%%%%%%%%%%%%%%%%%%%%%%%%%%%%%%%%%%%%%%%%%%%
% BEFORE SUBMISSION, PASTE THE .bbl FILE HERE,
% AND COMMENT OUT \bibliography{...} COMMAND

%\bibliography{droplet}

\begin{thebibliography}{27}
\expandafter\ifx\csname natexlab\endcsname\relax\def\natexlab#1{#1}\fi
\expandafter\ifx\csname bibnamefont\endcsname\relax
  \def\bibnamefont#1{#1}\fi
\expandafter\ifx\csname bibfnamefont\endcsname\relax
  \def\bibfnamefont#1{#1}\fi
\expandafter\ifx\csname citenamefont\endcsname\relax
  \def\citenamefont#1{#1}\fi
\expandafter\ifx\csname url\endcsname\relax
  \def\url#1{\texttt{#1}}\fi
\expandafter\ifx\csname urlprefix\endcsname\relax\def\urlprefix{URL }\fi
\providecommand{\bibinfo}[2]{#2}
\providecommand{\eprint}[2][]{\url{#2}}

\bibitem[{\citenamefont{Plateau}(1863, 1864, 1865)}]{plateau}
\bibinfo{author}{\bibfnamefont{J.}~\bibnamefont{Plateau}},
  \bibinfo{journal}{Annual Report of the Board of Regents of the Smithsonian
  Inst., Washington D.C.}  (\bibinfo{year}{1863, 1864, 1865}).

\bibitem[{\citenamefont{Cohen et~al.}(1974)\citenamefont{Cohen, Plasil, and
  Swiatecki}}]{cohen}
\bibinfo{author}{\bibfnamefont{S.}~\bibnamefont{Cohen}},
  \bibinfo{author}{\bibfnamefont{F.}~\bibnamefont{Plasil}}, \bibnamefont{and}
  \bibinfo{author}{\bibfnamefont{W.~J.} \bibnamefont{Swiatecki}},
  \bibinfo{journal}{Ann. Phys. (N.Y.)} \textbf{\bibinfo{volume}{82}},
  \bibinfo{pages}{557} (\bibinfo{year}{1974}).

\bibitem[{\citenamefont{Gamow}(1930)}]{gamow}
\bibinfo{author}{\bibfnamefont{G.}~\bibnamefont{Gamow}},
  \bibinfo{journal}{Proc. Roy. Soc. A} \textbf{\bibinfo{volume}{126}},
  \bibinfo{pages}{632} (\bibinfo{year}{1930}).

\bibitem[{\citenamefont{Cardoso and Gualtieri}(2006)}]{cardoso}
\bibinfo{author}{\bibfnamefont{V.}~\bibnamefont{Cardoso}} \bibnamefont{and}
  \bibinfo{author}{\bibfnamefont{L.}~\bibnamefont{Gualtieri}},
  \bibinfo{journal}{Class. Quantum Grav.} \textbf{\bibinfo{volume}{23}},
  \bibinfo{pages}{7151} (\bibinfo{year}{2006}).

\bibitem[{\citenamefont{Cardoso and Dias}(2006)}]{cardosoprl06}
\bibinfo{author}{\bibfnamefont{V.}~\bibnamefont{Cardoso}} \bibnamefont{and}
  \bibinfo{author}{\bibfnamefont{O.~J.~C.} \bibnamefont{Dias}},
  \bibinfo{journal}{Phys. Rev. Lett.} \textbf{\bibinfo{volume}{96}},
  \bibinfo{pages}{181601} (\bibinfo{year}{2006}).

\bibitem[{\citenamefont{Schunck et~al.}(2007)\citenamefont{Schunck, Dudek, and
  Herskind}}]{schunck}
\bibinfo{author}{\bibfnamefont{N.}~\bibnamefont{Schunck}},
  \bibinfo{author}{\bibfnamefont{J.}~\bibnamefont{Dudek}}, \bibnamefont{and}
  \bibinfo{author}{\bibfnamefont{B.}~\bibnamefont{Herskind}},
  \bibinfo{journal}{Phys. Rev. C} \textbf{\bibinfo{volume}{75}},
  \bibinfo{pages}{54304} (\bibinfo{year}{2007}).

\bibitem[{epa()}]{epaps}
\bibinfo{note}{See http://www.nottingham.ac.uk/\~{}ppzlev/Droplet}

\bibitem[{\citenamefont{Brown and Scriven}(1980)}]{brown80}
\bibinfo{author}{\bibfnamefont{R.~A.} \bibnamefont{Brown}} \bibnamefont{and}
  \bibinfo{author}{\bibfnamefont{L.~E.} \bibnamefont{Scriven}},
  \bibinfo{journal}{Proc. R. Soc. Lond. A} \textbf{\bibinfo{volume}{371}},
  \bibinfo{pages}{331} (\bibinfo{year}{1980}).

\bibitem[{\citenamefont{Wang et~al.}(1986)\citenamefont{Wang, Trinh,
  Croonquist, and Elleman}}]{wang86}
\bibinfo{author}{\bibfnamefont{T.~G.} \bibnamefont{Wang}},
  \bibinfo{author}{\bibfnamefont{E.~H.} \bibnamefont{Trinh}},
  \bibinfo{author}{\bibfnamefont{A.~P.} \bibnamefont{Croonquist}},
  \bibnamefont{and} \bibinfo{author}{\bibfnamefont{D.~D.}
  \bibnamefont{Elleman}}, \bibinfo{journal}{Phys. Rev. Lett.}
  \textbf{\bibinfo{volume}{56}}, \bibinfo{pages}{452} (\bibinfo{year}{1986}).

\bibitem[{\citenamefont{Wang et~al.}(1994)\citenamefont{Wang, Anilkumar, Lee,
  and Lin}}]{wang94}
\bibinfo{author}{\bibfnamefont{T.~G.} \bibnamefont{Wang}},
  \bibinfo{author}{\bibfnamefont{A.~V.} \bibnamefont{Anilkumar}},
  \bibinfo{author}{\bibfnamefont{C.~P.} \bibnamefont{Lee}}, \bibnamefont{and}
  \bibinfo{author}{\bibfnamefont{K.~C.} \bibnamefont{Lin}},
  \bibinfo{journal}{J. Fluid Mech.} \textbf{\bibinfo{volume}{276}},
  \bibinfo{pages}{389} (\bibinfo{year}{1994}).

\bibitem[{\citenamefont{Aussillous and Qu\'{e}r\'{e}}(2001)}]{aussillous01}
\bibinfo{author}{\bibfnamefont{P.}~\bibnamefont{Aussillous}} \bibnamefont{and}
  \bibinfo{author}{\bibfnamefont{D.}~\bibnamefont{Qu\'{e}r\'{e}}},
  \bibinfo{journal}{Nature} \textbf{\bibinfo{volume}{411}},
  \bibinfo{pages}{924} (\bibinfo{year}{2001}).

\bibitem[{\citenamefont{Chandrasekhar}(1965)}]{chandra65}
\bibinfo{author}{\bibfnamefont{S.}~\bibnamefont{Chandrasekhar}},
  \bibinfo{journal}{Proc. R. Soc. Lond. A} \textbf{\bibinfo{volume}{286}},
  \bibinfo{pages}{1} (\bibinfo{year}{1965}).

\bibitem[{\citenamefont{Ohsaka and Trinh}(2000)}]{ohsaka00}
\bibinfo{author}{\bibfnamefont{K.}~\bibnamefont{Ohsaka}} \bibnamefont{and}
  \bibinfo{author}{\bibfnamefont{E.~H.} \bibnamefont{Trinh}},
  \bibinfo{journal}{Phys. Rev. Lett.} \textbf{\bibinfo{volume}{84}},
  \bibinfo{pages}{1700} (\bibinfo{year}{2000}).

\bibitem[{\citenamefont{Beaugnon and
  Tournier}(1991{\natexlab{a}})}]{beaugnonNat}
\bibinfo{author}{\bibfnamefont{E.}~\bibnamefont{Beaugnon}} \bibnamefont{and}
  \bibinfo{author}{\bibfnamefont{R.}~\bibnamefont{Tournier}},
  \bibinfo{journal}{Nature} \textbf{\bibinfo{volume}{349}},
  \bibinfo{pages}{470} (\bibinfo{year}{1991}{\natexlab{a}}).

\bibitem[{\citenamefont{Beaugnon and
  Tournier}(1991{\natexlab{b}})}]{beaugnon91}
\bibinfo{author}{\bibfnamefont{E.}~\bibnamefont{Beaugnon}} \bibnamefont{and}
  \bibinfo{author}{\bibfnamefont{R.}~\bibnamefont{Tournier}},
  \bibinfo{journal}{J.Phys. {III} France} \textbf{\bibinfo{volume}{1}},
  \bibinfo{pages}{1423} (\bibinfo{year}{1991}{\natexlab{b}}).

\bibitem[{\citenamefont{Beaugnon et~al.}(2001)\citenamefont{Beaugnon, Fabregue,
  Billy, Nappa, and Tournier}}]{beaugnon01}
\bibinfo{author}{\bibfnamefont{E.}~\bibnamefont{Beaugnon}},
  \bibinfo{author}{\bibfnamefont{D.}~\bibnamefont{Fabregue}},
  \bibinfo{author}{\bibfnamefont{D.}~\bibnamefont{Billy}},
  \bibinfo{author}{\bibfnamefont{J.}~\bibnamefont{Nappa}}, \bibnamefont{and}
  \bibinfo{author}{\bibfnamefont{R.}~\bibnamefont{Tournier}},
  \bibinfo{journal}{Physica B} \textbf{\bibinfo{volume}{294-295}},
  \bibinfo{pages}{715} (\bibinfo{year}{2001}).

\bibitem[{\citenamefont{Berry and Geim}(1997)}]{berry}
\bibinfo{author}{\bibfnamefont{M.~V.} \bibnamefont{Berry}} \bibnamefont{and}
  \bibinfo{author}{\bibfnamefont{A.~K.} \bibnamefont{Geim}},
  \bibinfo{journal}{Eur. J. Phys.} \textbf{\bibinfo{volume}{18}},
  \bibinfo{pages}{307} (\bibinfo{year}{1997}).

\bibitem[{\citenamefont{Sueda et~al.}(2007)\citenamefont{Sueda, Katsuki,
  Nonomura, Kobayashi, and Tanimoto}}]{sueda}
\bibinfo{author}{\bibfnamefont{M.}~\bibnamefont{Sueda}},
  \bibinfo{author}{\bibfnamefont{A.}~\bibnamefont{Katsuki}},
  \bibinfo{author}{\bibfnamefont{M.}~\bibnamefont{Nonomura}},
  \bibinfo{author}{\bibfnamefont{R.}~\bibnamefont{Kobayashi}},
  \bibnamefont{and} \bibinfo{author}{\bibfnamefont{Y.}~\bibnamefont{Tanimoto}},
  \bibinfo{journal}{J. Phys. Chem. C} \textbf{\bibinfo{volume}{111}},
  \bibinfo{pages}{14389} (\bibinfo{year}{2007}).

\bibitem[{\citenamefont{Busse}(1984)}]{busse}
\bibinfo{author}{\bibfnamefont{F.~H.} \bibnamefont{Busse}},
  \bibinfo{journal}{J. Fluid Mech.} \textbf{\bibinfo{volume}{142}},
  \bibinfo{pages}{1} (\bibinfo{year}{1984}).

\bibitem[{\citenamefont{Wang}(1988)}]{wangreview}
\bibinfo{author}{\bibfnamefont{T.~G.} \bibnamefont{Wang}},
  \bibinfo{journal}{Adv. Appl. Mech.} \textbf{\bibinfo{volume}{26}},
  \bibinfo{pages}{1} (\bibinfo{year}{1988}).

\bibitem[{\citenamefont{Trinh and Wang}(1982)}]{trinh82}
\bibinfo{author}{\bibfnamefont{E.~H.} \bibnamefont{Trinh}} \bibnamefont{and}
  \bibinfo{author}{\bibfnamefont{T.~G.} \bibnamefont{Wang}},
  \bibinfo{journal}{J. Fluid Mech.} \textbf{\bibinfo{volume}{122}},
  \bibinfo{pages}{315} (\bibinfo{year}{1982}).

\bibitem[{\citenamefont{Trinh et~al.}(1988)\citenamefont{Trinh, Marston, and
  Robey}}]{trinh88}
\bibinfo{author}{\bibfnamefont{E.~H.} \bibnamefont{Trinh}},
  \bibinfo{author}{\bibfnamefont{P.~L.} \bibnamefont{Marston}},
  \bibnamefont{and} \bibinfo{author}{\bibfnamefont{J.~L.} \bibnamefont{Robey}},
  \bibinfo{journal}{J. Coll. Int. Sci.} \textbf{\bibinfo{volume}{124}},
  \bibinfo{pages}{95} (\bibinfo{year}{1988}).

\bibitem[{\citenamefont{Radyakin}(1979)}]{radyakin}
\bibinfo{author}{\bibfnamefont{N.~K.} \bibnamefont{Radyakin}},
  \bibinfo{journal}{Fluid Dyn., Springer NY} \textbf{\bibinfo{volume}{14}},
  \bibinfo{pages}{535} (\bibinfo{year}{1979}).

\bibitem[{\citenamefont{Ludu and Draayer}(1998)}]{ludu}
\bibinfo{author}{\bibfnamefont{A.}~\bibnamefont{Ludu}} \bibnamefont{and}
  \bibinfo{author}{\bibfnamefont{J.~P.} \bibnamefont{Draayer}},
  \bibinfo{journal}{Phys. Rev. Lett.} \textbf{\bibinfo{volume}{80}},
  \bibinfo{pages}{2125} (\bibinfo{year}{1998}).

\bibitem[{\citenamefont{Eriguchi and Hachisu}(1982)}]{eriguchi82}
\bibinfo{author}{\bibfnamefont{Y.}~\bibnamefont{Eriguchi}} \bibnamefont{and}
  \bibinfo{author}{\bibfnamefont{I.}~\bibnamefont{Hachisu}},
  \bibinfo{journal}{Prog. Theor. Phys.} \textbf{\bibinfo{volume}{67}},
  \bibinfo{pages}{844} (\bibinfo{year}{1982}).

\bibitem[{\citenamefont{Lacerda and Jewitt}(2007)}]{lacerda}
\bibinfo{author}{\bibfnamefont{P.}~\bibnamefont{Lacerda}} \bibnamefont{and}
  \bibinfo{author}{\bibfnamefont{D.~C.} \bibnamefont{Jewitt}},
  \bibinfo{journal}{Astron. J.} \textbf{\bibinfo{volume}{133}},
  \bibinfo{pages}{1393} (\bibinfo{year}{2007}).

\bibitem[{\citenamefont{Smarr}(1973)}]{smarr}
\bibinfo{author}{\bibfnamefont{L.}~\bibnamefont{Smarr}},
  \bibinfo{journal}{Phys. Rev. Lett.} \textbf{\bibinfo{volume}{30}},
  \bibinfo{pages}{71} (\bibinfo{year}{1973}).

\end{thebibliography}

%%%%%%%%%%%%%%%%%%%%%%%%%%%%%%%%%%%%%%%%%%%%%%%%%%%%%%%%%%%%%%%%%%%%%%%%%%%%%

\end{document}